\documentclass[%
 reprint,
 amsmath,amssymb,
 aps,
]{revtex4-1}

\usepackage{graphicx}
\usepackage{dcolumn}
\usepackage{bm}

\begin{document}

\preprint{APS/123-QED}

\title{$CP$ violation for $B^+_{c}\rightarrow D_{(s)}^+\pi^+\pi^-$ in Perturbative QCD}

\author{Gang L\"{u}}
 \email{ganglv66@sina.com}
\affiliation{%
 College of Science, Henan University of Technology, Zhengzhou 450001, China\\
 }%
 \author{Sheng-Tao Li}
 \email{394833074@qq.com}
\affiliation{%
 College of Science, Henan University of Technology, Zhengzhou 450001, China\\
 }%
 \author{Yu-Ting Wang}
 \email{1206166292@qq.com}
\affiliation{%
College of Science, Henan University of Technology, Zhengzhou 450001, China\\
 }


\begin{abstract}
In the perturbative QCD (PQCD) approach we study the direct $CP$ violation in
$B^+_{c}\rightarrow D_{(s)}^+\rho^0(\omega) \rightarrow D_{(s)}^+\pi^+\pi^-$
via the $\rho-\omega$ mixing mechanism. We find that the $CP$ violation can be
enhanced by $\rho-\omega$ mixing when the invariant masses of the $\pi^+\pi^-$ pairs are in the vicinity of the $\omega$
resonance. For the decay process $B^+_{c}\rightarrow D^+\rho^0(\omega) \rightarrow D^+\pi^+\pi^-$, the maximum $CP$ violation
can reach 7.5 {\%}.\\
\begin{description}
\item[PACS numbers]
{{13.25.Hw}, {11.30.Er}}
\end{description}
\end{abstract}

\maketitle
\section{\label{intro}Introduction}

$CP$ violation is an important topic in particle physics. Within the Standard Model (SM), $CP$ violation is originated from the weak phase in the Cabibbo--Kobayashi--Maskawa (CKM) matrix, along with the strong phase which usually arises from strong interactions \cite{cab,kob}. In the past few years more attention has been focused on the decays of $B$
meson system both theoretically and experimentally. Some events of $B_{c}$ mesons have been observed at Tevatron. Fortunately, a great number of events will appear at LHC in the foreseeable future. Recently, the LHCb Collaboration focused on three-body decays channels of $B^\pm\to \pi^\pm \pi^+\pi^-$ and $B^\pm\to K^\pm \pi^+\pi^-$  to probe large $CP$ violation \cite{J.M.,R.A.1,R.A.2}. The intriguing discoveries present us opportunities to detect $CP$ violation mechanism. The research of three-body decays of $B_c$ meson may be next topic for the LHCb experiments in the following years. In this paper, we focus on the interference from intermediate $\rho$ and $\omega$ mesons in the $B_{c}$ meson decays.

There has been remarkable progress in the study of exclusive $\overline{B_d^0} \to h_1 h_2$ and $B^\pm \to h_1 h_2$
decays, where $h_1, h_2$ are light pseudo-scalar and/or vector mesons. Historically, these decays were calculated in the
so-called naive factorization approach~\cite{bauer}, which was improved by including perturbative QCD
contributions~\cite{Ali:1997nh,9804363}. Currently, there are three popular theoretical approaches to study the dynamics of
these decays, which go under the name QCD factorization (QCDF)~\cite{qcdf}, perturbative QCD (PQCD)~\cite{pqcd,pqcd1}, and soft-collinear
effective theory (SCET)~\cite{scet}. All three are based on power expansion in $1/m_b$, where $m_b$ is the $b$-quark mass. Factorization of the hadronic matrix elements $\langle h_1 h_2 |{\cal O}_i|B\rangle$, where ${\cal O}_i$ is typically a four-quark or a magnetic moment type operator,
is shown to exist in the leading power in $1/m_b$ in a class of decays. But, these methods are different significantly due to the collinear degree
or transverse momenta. The power counting is different from the hard kernels
between QCDF and PQCD. It is important to extract the strong phase difference
for $CP$ violation. The more different feature of QCDF and PQCD is
the strong interaction scale at which of PQCD is low, typically of order
$1\sim 2 $ GeV, the case of QCDF is order $O(m_b)$ for the Wilson coefficients.

Direct $CP$ violating asymmetries in b-hadron decays occur through the interference of at least two amplitudes with the weak phase difference $\phi$ and the strong phase difference $\delta$. The weak phase difference is determined by the CKM matrix, while the strong phase is usually difficult to control. In order to acquire a large $CP$ violating asymmetries signal, we need to apply some phenomenological mechanism to obtain a large $\delta$. It has been shown that the charge symmetry violating mixing between $\rho$ and $\omega$ can be used to obtain a large strong phase difference which is required for large $CP$ violating asymmetries \cite{eno,gar,guo1,guo2,guo11,lei,gang1,gang2,gang3}. In this paper, we will investigate the $CP$ violation via $\rho-\omega$ mixing using PQCD approach in the decays of $B_c$ mesons.

In the perturbative QCD approach, at the rest frame of heavy $B$ meson, $B$ meson decays into two heavy quarks with large momenta. The hard interaction
dominants the decay amplitude from short distance due to not enough time to exchange soft gluons with final mesons. Since the final mesons
move very fast, a hard gluon kicks the light spectator quark of $B$ meson to form a fast moving final meson. Hence, the hard interaction is consist of
six quark operator. The non-perturbative dynamics are included in the meson wave function which can be extracted from experiment. The hard one can be
calculated by perturbation theory. Since $B_c$ meson has two heavy quarks (b and c quark), each of them can decay individually. It has been pointed out the c-quark decay processes can only produce about 1{\%} $CP$ violation \cite{guo11}. Hence, we only consider the contribution of b-quark decay in the processes that we are considering.

The remainder of this paper is organized as follows. In Sec.
\ref{sec:hamckm} we present the form of the effective Hamiltonian. In Sec. \ref{sec:cpv1} we give the calculating formalism of $CP$ violation
from $\rho-\omega$ mixing in $B^+_{c}\rightarrow D_{(s)}\pi^+\pi^-$ decay. We present the numerical results in Sec.
\ref{sec:numdis}. A summary and discussion are included in the
Sec. \ref{sec:conclusion}. The related functions defined in the text are given in Appendix.
\section{\label{sec:hamckm} The effective hamiltonian}
Based on the operator product expansion, the effective weak Hamiltonian can be expressed as \cite{buch}:

\begin{eqnarray}
{\cal H}_{eff} &=& {G_F\over \sqrt{2}} \{ \sum_{q=u,c} \xi_q [(C_1 (\mu)O^q_1 (\mu)   \nonumber \\
 &+& C_2 (\mu) O^q_2 (\mu) )+ \sum^{10}_{i=3} C_i (\mu) O_i (\mu) ] \},\;
\label{heff}
\vspace{2mm}
\end{eqnarray}
where $\xi_q$  = $V_{qp}V^*_{qb} (p=d, s)$ are CKM matrix elements, and $c_i(\mu)(i=1,2,..,10)$ are the Wilson coefficients, which are calculable in the renormalization group improved perturbation theory and are scale dependent. In the present case, we work with the renormalization scheme independent Wilson coefficients and use the values of the Wilson coefficients at the renormalization scale $\mu\approx m_b$. $G_F$ represents Fermi constant and $O_i$ is the effective four quark operator, which can be written as
\begin{eqnarray}
O^{u}_1&=& \bar d_\alpha \gamma_\mu(1-\gamma_5)u_\beta\bar
u_\beta\gamma^\mu(1-\gamma_5)b_\alpha,\nonumber\\
O^{u}_2&=& \bar d \gamma_\mu(1-\gamma_5)u\bar
u\gamma^\mu(1-\gamma_5)b,\nonumber\\
O_3&=& \bar d \gamma_\mu(1-\gamma_5)b \sum_{q'}
\bar q' \gamma^\mu(1-\gamma_5) q',\nonumber\\
O_4 &=& \bar d_\alpha \gamma_\mu(1-\gamma_5)b_\beta \sum_{q'}
\bar q'_\beta \gamma^\mu(1-\gamma_5) q'_\alpha,\nonumber\\
O_5&=&\bar d \gamma_\mu(1-\gamma_5)b \sum_{q'} \bar q'
\gamma^\mu(1+\gamma_5)q',\nonumber\\
O_6& = &\bar d_\alpha \gamma_\mu(1-\gamma_5)b_\beta \sum_{q'}
\bar q'_\beta \gamma^\mu(1+\gamma_5) q'_\alpha,\nonumber\\
O_7&=& \frac{3}{2}\bar d \gamma_\mu(1-\gamma_5)b \sum_{q'}
e_{q'}\bar q' \gamma^\mu(1+\gamma_5) q',\nonumber\\
O_8 &=&\frac{3}{2} \bar d_\alpha \gamma_\mu(1-\gamma_5)b_\beta \sum_{q'}
e_{q'}\bar q'_\beta \gamma^\mu(1+\gamma_5) q'_\alpha,\nonumber\\
O_9&=&\frac{3}{2}\bar d \gamma_\mu(1-\gamma_5)b \sum_{q'} e_{q'}\bar q'
\gamma^\mu(1-\gamma_5)q',\nonumber\\
O_{10}& = &\frac{3}{2}\bar d_\alpha \gamma_\mu(1-\gamma_5)b_\beta \sum_{q'}
e_{q'}\bar q'_\beta \gamma^\mu(1-\gamma_5) q'_\alpha,\nonumber\\
&&
\label{O1-10}
\vspace{2mm}
\end{eqnarray}
where $\alpha$ and $\beta$ are color indices, and $q'$ running through all the light flavour quarks. In Eq.(\ref{O1-10}) $O_1^u$ and $O_2^u$ are the tree level and QCD corrected operators, $O_3$--$O_6$ are QCD penguin operators and $O_7$--$O_{10}$ are
the operators associated with electroweak penguin diagrams.

In the PQCD approach, three scales are involved:
the W-boson mass $m_{W}$ associated weak interaction,
the hard scale $t$, and factorization scale $1/b$
($b$ is the conjugate variable of the parton transverse momenta $k_{T}$). The decay amplitude is then factorized into the convolution of the hard subamplitude, the Wilson coefficient and the Sudakov factor with the meson wave functions, all of which are well-defined and gauge invariant. Therefore, the three scale factorization formula for exclusive nonleptonic $B$ meson decays is then written as
\begin{eqnarray}
C(t)&\otimes& H(x,t)\otimes \Phi(x)\otimes exp[-s(P,b)   \nonumber \\
&-&2\int^{t}_{1/b}
\frac{d\mu}{\mu}\gamma_q(\alpha_{s}(\mu))],
\label{eq: gen-expression}
\end{eqnarray}
where $C(t)$ are the corresponding Wilson coefficients. $\Phi(x)$ are the meson wave functions and the variable t denotes the largest mass scale of hard process H that is six-quark effective theory. Sudakov factor coming from renormalization summation and threshold summation is introduced to solve the endpoint diverging. It can handle with endpoint diverging problem properly by introducing Sudakov factor \cite{npb193381}. The Sudakov evolution exp [-s(P,b)] are from the resummation of double logarithms  $ln^2 (Pb)$, with P denoting the dominant light-cone component of meson momentum. $\gamma_q=-\alpha_{s}/\pi$ is quark anomalous dimension in axial gauge.

\section{\label{sec:cpv1}$CP$ violation in $B^+_{c}\rightarrow D_{(s)}^+\rho^0(\omega) \rightarrow D_{(s)}^+\pi^+\pi^-$}
\subsection{\label{subsec:form}Formalism}

In the vector meson dominance model, the photon propagator is dressed by coupling to vector meson. Based on the same mechanism, $\rho-\omega$ mixing was proposed. According to the effective Hamiltonian, the amplitude $\mathcal{A}$ for $B^+_{c}\rightarrow D_{(s)}^+\pi^+\pi^-$ can be divided into two parts:
\begin{eqnarray}
\mathcal{A}=\big<D_{(s)}^+\pi^+\pi^-|H^T|B^+_{c}\big>+\big<D_{(s)}^+\pi^+\pi^-|H^P|B^+_{c}\big>,\label{A}\nonumber\\
\end{eqnarray}
with $H^T$ and $H^P$ being the Hamiltonian for the tree and
penguin operators, respectively.

We can define the relative magnitudes and phases between the tree
and penguin operator contributions as follows:
\begin{eqnarray}
\mathcal{A}=\big<D_{(s)}^+\pi^+\pi^-|H^T|B^+_{c}\big>[1+re^{i(\delta+\phi)}],\label{A'}
\end{eqnarray}
where $\delta$ and $\phi$ are strong and weak phases, respectively.
$\phi$ arises from the CP-violating phase in the CKM matrix, which
is arg$[V_{tb}V^{*}_{tq}/(V_{ub}V^{*}_{uq})](q=d,s)$. The parameter $r$ is the
absolute value of the ratio of penguin and tree amplitudes:
\begin{eqnarray}
r\equiv\Bigg|\frac{\big<D_{(s)}^+\pi^+\pi^-|H^P|B^+_{c}\big>}{\big<D_{(s)}^+\pi^+\pi^-|H^T|B^+_{c}\big>}\Bigg|
\label{r}.
\end{eqnarray}
The amplitude for $B^-_{c} \rightarrow {D}^-_{(s)}\pi^+\pi^-$ is
\begin{eqnarray}
\bar{\mathcal{A}}&=&\big<{D}^-_{(s)}\pi^+\pi^-|H^T|{B^-_{c}}\big>+\big<{D}^-_{(s)}\pi^+\pi^-|H^P|{B^-_{c}}\big>. \label{asy}   \nonumber\\
\end{eqnarray}
In this work, we only consider $\rho^0$ and $\omega$ resonances. The $CP$ violating asymmetry for $B^+_{c}\rightarrow D^+_{(s)}\pi^+\pi^-$ is defined as
\begin{eqnarray}
A_{cp} \equiv\frac{|A|^2-|\bar{A}|^2}{|A|^2+|\bar{A}|^2}=\frac{-2r
{\rm{sin}}\delta {\rm{sin}}\phi}{1+2r {\rm{cos}}\delta
{\rm{cos}}\phi+r^2}. \label{acp}
\end{eqnarray}
From Equation (\ref{acp}), one can find the $CP$ violation
depends on the weak phase difference and the strong phase
difference. The weak phase is determined for a specific decay process.
Hence, in order to obtain a large $CP$ violation, we
need some mechanism to make sin$\delta$ large. It has been found
that $\rho-\omega$ mixing (which was proposed based on vector
meson dominance \cite{HB1997}) leads to
a large strong phase difference
\cite{gar,guo1,guo2,guo11,lei,gang1,gang2,gang3}. Based on $\rho-\omega$ mixing and working to the first order
of isospin violation, we have the following results:
\begin{eqnarray}
\big<D_{(s)}^+\pi^+\pi^-|H^T|B^+_{c}\big>=\frac{g_{\rho}}{s_{\rho}s_{\omega}}\widetilde{\Pi}_{\rho\omega}t_{\omega}+\frac{g_{\rho}}{s_{\rho}}t_{\rho},
\label{Htr}\\
\big<D_{(s)}^+\pi^+\pi^-|H^P|B^+_{c}\big>=\frac{g_{\rho}}{s_{\rho}s_{\omega}}\widetilde{\Pi}_{\rho\omega}p_{\omega}+\frac{g_{\rho}}{s_{\rho}}p_{\rho}.
\label{Hpe}
\end{eqnarray}
where $t_v$($v=\rho$ or $\omega$) is the tree amplitudes and $p_v$ is the penguin amplitudes for producing an intermediate vector meson V. $g_{\rho}$ is
the coupling for $\rho^0\rightarrow\pi^+\pi^-$;
$\widetilde{\Pi}_{\rho\omega}$ is the effective $\rho-\omega$
mixing amplitude which also effectively includes the direct
coupling $\omega\rightarrow\pi^+\pi^-$. $s_{V}$, $m_{V}$ and $\Gamma_V$($V$=$\rho$ or
$\omega$) is the inverse propagator, mass and decay rate of the vector meson $V$, respectively.
\begin{eqnarray}
s_V=s-m_V^2+{\rm{i}}m_V\Gamma_V.
\end{eqnarray}
with $\sqrt{s}$ being the invariant masses of the $\pi^+\pi^-$
pairs.

We stress that the direct coupling $\omega\rightarrow\pi^+\pi^-$ is
effectively absorbed into $\widetilde{\Pi}_{\rho\omega}$, leading to the
explicit $s$ dependence of
$\widetilde{\Pi}_{\rho\omega}$ \cite{oco}. However, the $s$ dependence
of $\widetilde{\Pi}_{\rho\omega}$ is negligible in practice. We can make the expansion
$\widetilde{\Pi}_{\rho\omega}(s)=\widetilde{\Pi}_{\rho\omega}(m_{\omega}^2)+(s-m_{\omega})\widetilde{\Pi}_{\rho\omega}^\prime(m_{\omega}^2)$. The $\rho-\omega$ mixing parameters were determined in the
fit of Gardner and O'Connell \cite{gard}:
\begin{eqnarray}
\mathfrak{Re}\widetilde{\Pi}_{\rho\omega}(m_{\omega}^2)&=&-3500\pm300
\rm{MeV}^2,\nonumber\\{\mathfrak{Im}}\widetilde{\Pi}_{\rho\omega}(m_{\omega}^2)&=&-300\pm300
\textrm{MeV}^2,\nonumber\\\widetilde{\Pi}_{\rho\omega}^\prime(m_{\omega}^2)&=&0.03\pm0.04.
\end{eqnarray}
From Eqs. (\ref{A})(\ref{A'})(\ref{Htr})(\ref{Hpe}) one has
\begin{eqnarray}
re^{i\delta}e^{i\phi}=\frac{\widetilde{\Pi}_{\rho\omega}p_{\omega}+s_{\omega}p_{\rho}}{\widetilde{\Pi}_{\rho\omega}t_{\omega}+s_{\omega}t_{\rho}},
\label{rdtdirive}
\end{eqnarray}
Defining
\begin{eqnarray}
\frac{p_{\omega}}{t_{\rho}}\equiv r^\prime
e^{i(\delta_q+\phi)},\quad\frac{t_{\omega}}{t_{\rho}}\equiv
\alpha
e^{i\delta_\alpha},\quad\frac{p_{\rho}}{p_{\omega}}\equiv
\beta e^{i\delta_\beta}, \label{def}
\end{eqnarray}
where $\delta_\alpha$, $\delta_\beta$ and $\delta_q$ are strong
phases. One finds the following expression from Eqs.
(\ref{rdtdirive})(\ref{def}):
\begin{eqnarray}
re^{i\delta}=r^\prime
e^{i\delta_q}\frac{\widetilde{\Pi}_{\rho\omega}+\beta
e^{i\delta_\beta}s_{\omega}}{\widetilde{\Pi}_{\rho\omega}\alpha
e^{i\delta_\alpha}+s_{\omega}}. \label{rdt}
\end{eqnarray}
In order to get the $CP$ violating asymmetry in Eq.
(\ref{acp}), sin$\phi$ and cos$\phi$ are needed. The weak phase $\phi$ is
fixed by the CKM matrix elements. In the Wolfenstein
parametrization \cite{wol}, one has
\begin{eqnarray}
{\rm sin}\phi &=&\frac{\eta}{\sqrt{[\rho(1-\rho)-\eta^2]^2+\eta^2}}, \nonumber \\
{\rm cos}\phi &=&\frac{\rho(1-\rho)-\eta^2}{\sqrt{[\rho(1-\rho)-\eta^2]^2+\eta^2}}.
\label{3l1}
\vspace{2mm}
\end{eqnarray}

\subsection{\label{subsec:cal}Calculational details}
From Equations (\ref{acp})(\ref{rdtdirive})(\ref{def}), in order to obtain
the formulas of the $CP$ violation, we calculate the
amplitudes $t_{\rho}$, $t_{\omega}$, $p_{\rho}$ and $p_{\omega}$ in PQCD approach,
which can be decomposed in terms of tree-level and penguin-level amplitudes due to
the CKM matrix elements of $V_{ud}V^{*}_{ub}$, $V_{us}V^{*}_{ub}$, $V_{td}V^{*}_{tb}$ and  $V_{ts}V^{*}_{tb}$.
In the following, we calculate the decay amplitudes for $B^+_{c}\rightarrow D^+\rho^{0}(\omega)$
and $B^+_{c}\rightarrow D_{s}^+\rho^{0}(\omega)$ which we will use in the next paragraph. The PQCD function of $F$ and $M$ can be found in appendix.

\subsubsection{\label{subsubsec:BcD}The decay amplitudes of $B^+_{c}\rightarrow D^+\rho^{0}(\omega)$}

With the Hamiltonian (\ref{heff}), depending on CKM matrix elements of $V_{ud}V^{*}_{ub}$ , $V_{us}V^{*}_{ub}$ , $V_{td}V^{*}_{tb}$ and  $V_{ts}V^{*}_{tb}$,
the decay amplitudes for $B^+_{c}\rightarrow D^+\rho^{0}$ in PQCD can be
written as
\begin{eqnarray}
\sqrt{2}\mathcal{A}(B^+_{c}\rightarrow D^+\rho^{0})=V_{ud}V^{*}_{ub}t_{\rho}+V_{td}V^{*}_{tb}p_{\rho} \label{BcDrho1}
\end{eqnarray}
where
\begin{eqnarray}
t_{\rho}&=&(C_{1}+\frac{1}{3}C_{2}) F_{e}^{LL}+C_{2} M_{e}^{LL}   \nonumber\\
&+&(C_{2}+\frac{1}{3}C_{1}) F_{a}^{LL}+C_{1} M_{a}^{LL} \label{trho1}
\end{eqnarray}
and
\begin{eqnarray}
P_{\rho}&=&(C_{2}+\frac{1}{3}C_{1}) F_{a}^{LL}+C_{1} M_{a}^{LL}  \nonumber\\
&-&[(\frac{3}{2}C_{10}-C_{3}+\frac{1}{2}C_{9}) M_{e}^{LL}  \nonumber\\
&-&(C_{3}+C_{9}) M_{a}^{LL}+(-C_{5} + \frac{1}{2}C_{7})M_{e}^{LR}     \nonumber\\
&+& (-C_{4}-\frac{1}{3}C_{3}-C_{10}-\frac{1}{3}C_{9} )F_{a}^{LL}   \nonumber\\
&+&(C_{10}+\frac{5}{3}C_{9}-\frac{1}{3}C_{3} -C_{4}-\frac{3}{2}C_{7}-\frac{1}{2}C_{8}) F_{e}^{LL}    \nonumber\\
&+&(-C_{6}-\frac{1}{3}C_{5}+\frac{1}{2}C_{8} +\frac{1}{6}C_{7})F_{e}^{SP}      \nonumber\\
&-&(C_{5}+C_{7})M_{a}^{LR}    \nonumber\\
&+&(-C_{6}-\frac{1}{3}C_{5}-C_{8}-\frac{1}{3}C_{7})F_{a}^{SP}]  \label{Prho1}
\end{eqnarray}
The decay amplitude for $B^+_{c}\rightarrow D^+\omega$ can be written as
\begin{eqnarray}
\sqrt{2}\mathcal{A}(B^+_{c}\rightarrow D^+\omega)=V_{ud}V^{*}_{ub}t_{\omega}-V_{td}V^{*}_{tb}p_{\omega}, \label{BcDomega1}
\end{eqnarray}
where
\begin{eqnarray}
t_{\omega}&=&(C_{1}+\frac{1}{3}C_{2}) F_{e}^{LL}+C_{2} M_{e}^{LL}   \nonumber\\
&-&[(C_{2}+\frac{1}{3}C_{1}) F_{a}^{LL}+C_{1} M_{a}^{LL}] \label{tomega1}
\end{eqnarray}
and
\begin{eqnarray}
p_{\omega}&=&(C_{2}+\frac{1}{3}C_{1}) F_{a}^{LL}+C_{1} M_{a}^{LL}  \nonumber\\
&+&[(2C_{4}+C_{3}+\frac{1}{2}C_{10} -\frac{1}{2}C_{9})M^{LL}_{e}  \nonumber\\
&+&(C_{3}+C_{9})M^{LL}_{e}   \nonumber\\
&+&(C_{5}-\frac{1}{2}C_{7})M^{LR}_{e}+(C_{5}+C_{7})M^{LR}_{a}   \nonumber\\
&+&(C_{4}+\frac{1}{3}C_{3}+C_{10}+\frac{1}{3}C_{9})F^{LL}_{a}   \nonumber\\
&+&(\frac{7}{3}C_{3}+\frac{5}{3}C_{4}+\frac{1}{3}(C_{9}-C_{10}))F^{LL}_{e}  \nonumber\\
&+&(2C_5+\frac{2}{3}C_{6}+\frac{1}{2}C_{7}+\frac{1}{6}C_{8})F^{LR}_{e}     \nonumber\\
&+&(C_{6}+\frac{1}{3}C_{5}-\frac{1}{2}C_{8}-\frac{1}{6}C_{7})F^{SP}_{e}    \nonumber\\
&+&(C_{6}+\frac{1}{3}C_{5}+C_{8}+\frac{1}{3}C_{7})F^{SP}_{a}] \label{pomega1}
\end{eqnarray}
Based on the definition of (\ref{def}), we can get
\begin{eqnarray}
\alpha e^{i\delta_\alpha}&=&\frac{t_{\omega}}{t_{\rho}}, \label{eq:afaform} \\
\beta e^{i\delta_\beta}&=&\frac{p_{\rho}}{p_{\omega}}, \label{eq:btaform}\\
r^\prime e^{i\delta_q}&=&\frac{p_{\omega}}{t_{\rho}}
\times\bigg|\frac{V_{tb}V_{td}^*}{V_{ub}V_{ud}^*}\bigg|,  \label{eq:delform}
\end{eqnarray}
where
\begin{equation}
\left|\frac{V_{tb}V^{*}_{td}}{V_{ub}V^{*}_{ud}}\right|=\frac{\sqrt{[\rho(1-\rho)-\eta^2]^2+\eta^2}}{(1-\lambda^2/2)(\rho^2+\eta^2)}
\label{3p}
\vspace{2mm}
\end{equation}

\subsubsection{\label{subsubsec:BcDs}The decay amplitudes of $B^+_{c}\rightarrow D^+_{s}\rho^{0}(\omega)$}

The decay amplitudes for $B^+_{c}\rightarrow D^+_{s}\rho^{0}$ can be
written as
\begin{eqnarray}
\sqrt{2}\mathcal{A}(B^+_{c}\rightarrow D^+_{s}\rho^{0})=V_{us}V^{*}_{ub}t_{\rho}-V_{ts}V^{*}_{tb}p_{\rho}, \label{BcDsrho1}
\end{eqnarray}
where
\begin{eqnarray}
t_{\rho}&=&(C_{1}+\frac{1}{3}C_{2}) F_{e}^{LL}+C_{2} M_{e}^{LL} \label{trho2}
\end{eqnarray}
and
\begin{eqnarray}
p_{\rho}&=&[(\frac{1}{2}(3C_9 +C_{10})F^{LL}_e +\frac{1}{2}(3C_7 +C_{8}))F^{LR}_e   \nonumber \\
&+& \frac{3}{2}C_{10}M^{LL}_e +\frac{3}{2}C_8 M^{SP}_e]  \label{Prho2}
\end{eqnarray}

The decay amplitudes for $B^+_c\rightarrow D^+_{s}\omega$ can be
written as
\begin{eqnarray}
\sqrt{2}\mathcal{A}(B^+_c\rightarrow D^+_{s}\omega)=V_{us}V^{*}_{ub}t_{\omega}-V_{ts}V^{*}_{tb}p_{\omega}, \label{BcDsomega1}
\end{eqnarray}
where
\begin{eqnarray}
t_{\omega}&=&(C_{1}+\frac{1}{3}C_{2}) F_{e}^{LL}+C_{2} M_{e}^{LL}  \label{tomega2}
\end{eqnarray}
and
\begin{eqnarray}
p_{\omega}&=&[(2C_4+\frac{1}{2}C_{10})M^{LL}_e +(2C_6+\frac{1}{2}C_8)M^{SP}_e   \nonumber \\
&+&(2C_3 +\frac{2}{3}C_4 +\frac{1}{2}C_9 +\frac{1}{6}C_{10} )F^{LL}_e  \nonumber \\
&+&(2C_5 +\frac{2}{3}C_6 +\frac{1}{2}C_7+\frac{1}{6}C_8)F^{LR}_e]   \label{Pomega2}
\end{eqnarray}
Similarity, we can also obtain the strong phase from the Eqs. (\ref{eq:afaform})(\ref{eq:btaform})(\ref{eq:delform})(\ref{3p}).

\section{\label{sec:numdis}Numerical results}
\subsection{\label{int}Input parameters}
In the numerical calculations, we have several parameters. The Wilson coefficients, $C_i(\mu)$, take the following values \cite{pqcd1}:
\begin{eqnarray}
C_1 &=&-0.2703, \;\; \;C_2=1.1188,\nonumber\\
C_3 &=& 0.0126,\;\;\;C_4 = -0.0270,\nonumber\\
C_5 &=& 0.0085,\;\;\;C_6 = -0.0326,\nonumber\\
C_7 &=& 0.0011,\;\;\;C_8 = 0.0004,\nonumber\\
C_9&=& -0.0090,\;\;\;C_{10} = 0.0022,
\label{2k}
\vspace{2mm}
\end{eqnarray}

The CKM matrix, which should be determined from experiments, can
be expressed in terms of the Wolfenstein parameters,
$A,\lambda,\rho$ and $\eta$ \cite{wol}:
\begin{eqnarray}
\left(\begin{array}{ccc}1-\frac{1}{2}\lambda^2 & \lambda &
A\lambda^3(\rho-i\eta)\\-\lambda & 1-\frac{1}{2}\lambda^2 &
A\lambda^2\\ A\lambda^3(1-\rho-i\eta) & -A\lambda^2 & 1
\end{array}\right),
\end{eqnarray}
where $O(\lambda^4)$ corrections are neglected. The latest values
for the parameters in the CKM matrix are \cite{npp37}:
\begin{eqnarray}
&& \lambda=0.2253\pm0.0007,\quad A=0.808_{-0.015}^{+0.022},\nonumber \\
&& \bar{\rho}=0.132_{-0.014}^{+0.022},\quad
\bar{\eta}=0.341\pm0.013,\label{eq: rhobarvalue}
\end{eqnarray}
with
\begin{eqnarray}
 \bar{\rho}=\rho(1-\frac{\lambda^2}{2}),\quad
\bar{\eta}=\eta(1-\frac{\lambda^2}{2}).\label{eq: rho rhobar
relation}
\end{eqnarray}
From Eqs. (\ref{eq: rhobarvalue}) ( \ref{eq: rho rhobar relation})
we have
\begin{eqnarray}
0.121<\rho<0.158,\quad  0.336<\eta<0.363.\label{eq: rho value}
\end{eqnarray}
We adopt the results from \cite{npp37}:
\begin{eqnarray}
|V_{ub}|&=&(3.89\pm0.44) \times {10}^{-3}, \hspace{1cm} |V_{ud}|=0.97425,   \nonumber \\
|V_{cb}|&=&0.0406, \hspace{3.18cm}  |V_{cd}|=0.23   \nonumber \\
|V_{us}|&=&0.2252, \hspace{3.2cm}   |V_{cs}|=1.023,        \nonumber \\
m_{b}&=&4.2GeV,    \hspace{3.25cm}   m^{\pi}_{0}=1.4GeV,    \nonumber  \\
m^{K}_{0}&=&1.6GeV,  \hspace{3.17cm}  m^{{\eta}_{q}}_{0}=1.07GeV,  \nonumber  \\
\gamma &=&({73}^{+22}_{-25})^{\circ}  \hspace{3.29cm}  m_{c}=1.27GeV, \nonumber  \\
m^{{\eta}_{s}}_{0}&=&1.92GeV,   \hspace{2.72cm}   \Lambda_{QCD}^{5}=0.112GeV.    \nonumber \\    \label{PDG2010}
\end{eqnarray}

\subsection{\label{subsec:NumeA} $CP$ violation in $B^+_{c}\rightarrow D_{(s)}^+\pi^+\pi^-$}
In the numerical results, we find that
the $CP$ violation can be enhanced via  $\rho-\omega$ mixing for the decay channel $B^+_{c}\rightarrow D_{(s)}^+\pi^+\pi^-$ when the
invariant mass of $\pi^{+}\pi^{-}$ is in the vicinity of the
$\omega$ resonance. The $CP$ violation depends on the weak phase difference from CKM matrix elements and the strong phase difference which is difficult to control.
The CKM matrix elements, which relate to
$\rho$, $\eta$, $\lambda$ and $A$,
are given in Eq.(\ref{eq: rhobarvalue}). The uncertainties due to the CKM matrix elements come from $\rho$, $\eta$, $\lambda$ and $A$. In our numerical
calculations, we let $\rho$, $\eta$, $\lambda$ and $A$ vary among the limiting values.  The numerical results are shown from Fig.1 to Fig.6 with the different parameter values of CKM matrix elements. The dash line, dot line and solid line corresponds to the maximum, middle, and minimum CKM matrix element for the decay channel of $B^+_{c}\rightarrow D^+_{(s)}\pi^+\pi^{-}$, respectively. In Fig.1 and Fig.2, we
give the central value of $CP$ violating asymmetry
as a function of $\sqrt{s}$. From the Fig.1 and Fig.2 one can see
the $CP$ violation parameter is dependent on $\sqrt{s}$
and changes rapidly due to $\rho-\omega$ mixing when the
invariant mass of $\pi^{+}\pi^{-}$ is in the vicinity of the
$\omega$ resonance. We can see that the CP violating asymmetry vary
from around $2\%$ to around $7.5\%$ for the decay channel of $B_c\rightarrow D^{+}\pi^+\pi^{-}$
when $\sqrt{s}=0.786 GeV$ in Fig.1. As can be seen from Fig.2 the CP violating asymmetry vary
from around $5.3\%$ to around $7.2\%$ for the decay channel of $B_c\rightarrow D^{+}_s\pi^+\pi^{-}$
when $\sqrt{s}=0.774 GeV$.

\begin{figure}
\includegraphics[width=0.48\textwidth]{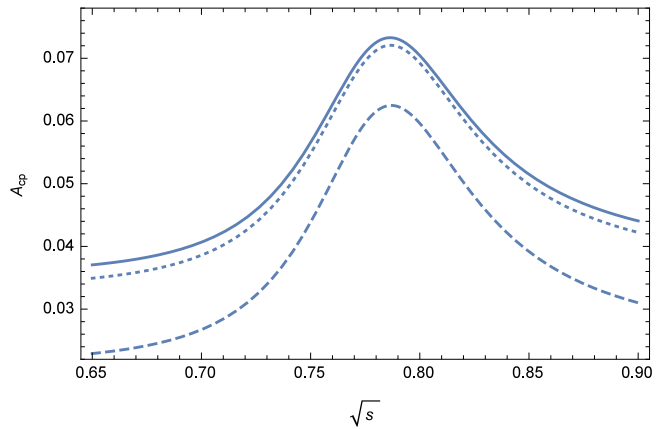}\\
\caption{The $CP$ violating asymmetry, $A_{cp}$, as a function of $\sqrt{s}$ for
different CKM matrix elements. The dash line, dot line and solid line corresponds to the maximum, middle, and minimum CKM matrix element for the decay channel of $B^+_{c}\rightarrow D^+\pi^+\pi^{-}$, respectively.}\label{Acp1 plot}
\end{figure}

\begin{figure}
\includegraphics[width=0.48\textwidth]{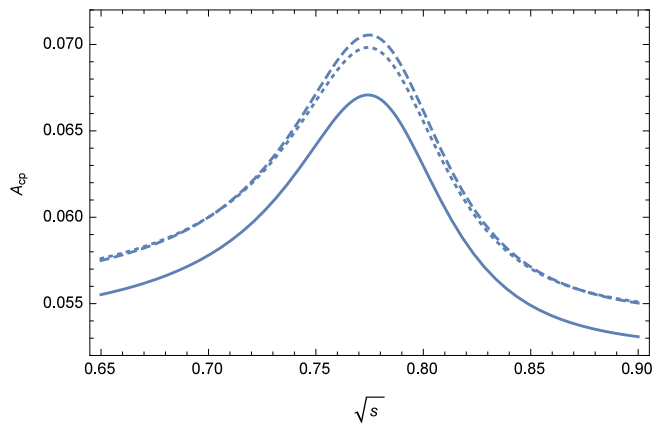}\\
\caption{The $CP$ violating asymmetry, $A_{cp}$, as a function of $\sqrt{s}$ for
different CKM matrix elements. The dash line, dot line and solid line corresponds to the maximum, middle, and minimum CKM matrix element for the decay channel of $B^+_{c}\rightarrow D^+_s\pi^+\pi^{-}$, respectively.}\label{Acp2 plot}
\end{figure}
From Eq.(\ref{acp}), we can see that the $CP$ violating parameter is related to
sin$\delta$ and $r$. The plots of
$\sin\delta$ and $r$ as a function of $\sqrt{s}$
 are shown in Fig.3, Fig.4, Fig.5 and Fig.6, respectively. It
can be seen that $\sin\delta$ and $r$ change sharply at the range of $\rho-\omega$ resonance. In Fig.3 and Fig.4, we show the plot of $\sin\delta$
as a function of $\sqrt{s}$. We can see that the $\rho-\omega$ mixing mechanism leads to the strong phase at the $\omega$ resonance for
the processes of $B^+_c\rightarrow D^{+}\pi^+\pi^{-}$ and
$B^+_c\rightarrow D^{+}_s\pi^+\pi^{-}$. One can
find $\rho-\omega$ mixing make the $\sin\delta$ value oscillate from
$-0.03$ to $-0.095$ and $0.118$ to $0.223$ for the decay processes of $B^+_c\rightarrow D^{+}\pi^+\pi^{-}$ and
$B^+_c\rightarrow D^{+}_s\pi^+\pi^{-}$, respectively. From Fig.5 and Fig.6, one can see that $r$ increases slowly for the channel of $B^+_{c}\rightarrow D^+_{(s)}\pi^+\pi^{-}$ when the invariant masses of the $\pi^+\pi^-$
pairs are in the vicinity of the $\omega$ resonance.

\begin{figure}
\includegraphics[width=0.48\textwidth]{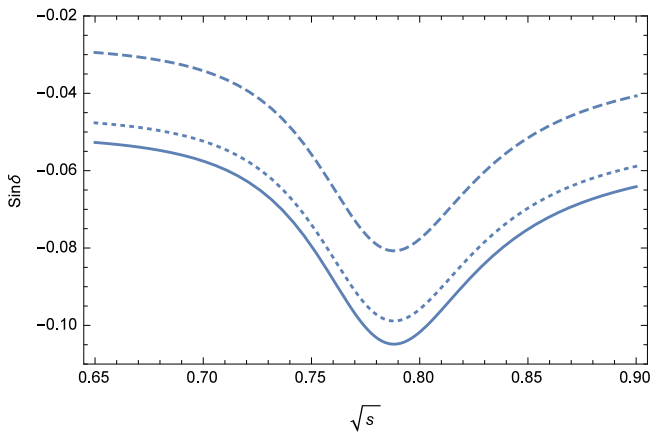}\\
\caption{$sin\delta$ as a function of $\sqrt{s}$ for
different CKM matrix elements. The dash line, dot line and solid line corresponds to the maximum, middle, and minimum CKM matrix element for the decay channel of $B^+_{c}\rightarrow D^+\pi^+\pi^{-}$, respectively.}\label{sin1 plot}
\end{figure}

\begin{figure}
\includegraphics[width=0.48\textwidth]{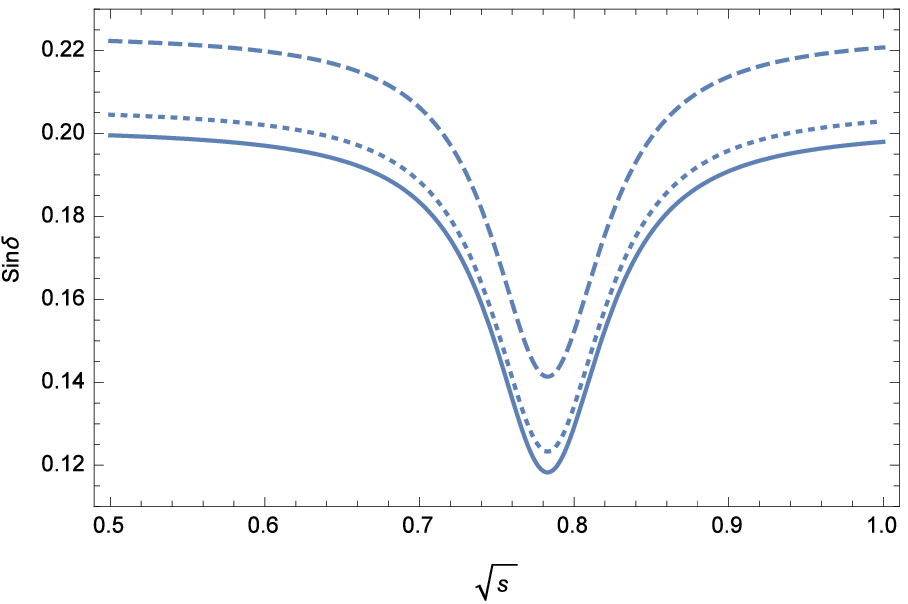}\\
\caption{$sin\delta$ as a function of $\sqrt{s}$ for
different CKM matrix elements. The dash line, dot line and solid line corresponds to the maximum, middle, and minimum CKM matrix element for the decay channel of $B^+_{c}\rightarrow D^+_s\pi^+\pi^{-}$, respectively.}\label{sin2 plot}
\end{figure}

\begin{figure}
\includegraphics[width=0.48\textwidth]{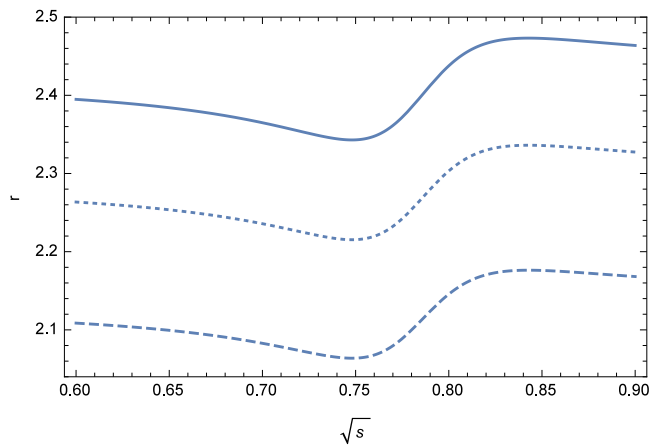}\\
\caption{$r$ as a function of $\sqrt{s}$ for
different CKM matrix elements. The dash line, dot line and solid line corresponds to the maximum, middle, and minimum CKM matrix element for the decay channel of $B^+_{c}\rightarrow D^+\pi^+\pi^{-}$, respectively.}\label{r1 plot}
\end{figure}

For the processes of above decay channels, $\rho-\omega$ mixing does enhance $CP$
violating asymmetries and provide a mechanism for producing large $CP$
violation in perturbatibe QCD. Meanwhile we find $\rho-\omega$ mixing presents strong phase so as to make sin$\delta$ big and can also change the value of $r$. However, we find that the effect of the change of $r$ on $A_{cp}$ is small compared with the
case of sin$\delta$ for the processes we are considering.

\section{\label{sec:conclusion}Summary and Discussion}
In this paper, we have studied the $CP$ violation in the decay of
$B^+_{c}\rightarrow D_{(s)}^+\rho^0(\omega) \rightarrow D_{(s)}^+\pi^+\pi^-$
due to the contribution of $\rho-\omega$ mixing in PQCD approach.
It is found that $\rho-\omega$ mixing
can cause a large strong phase difference so that
large $CP$ violation can be obtained at the $\omega$ resonance.
As a result, it is found that
the maximum $CP$ violation can reach $7.5\%$.

\begin{figure}
\includegraphics[width=0.48\textwidth]{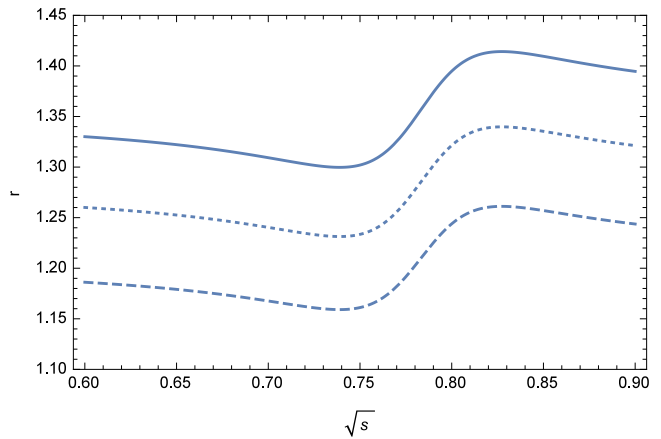}\\
\caption{$r$ as a function of $\sqrt{s}$ for
different CKM matrix elements. The dash line, dot line and solid line corresponds to the maximum, middle, and minimum CKM matrix element for the decay channel of $B^+_{c}\rightarrow D^+_s\pi^+\pi^{-}$, respectively.}\label{r2 plot}
\end{figure}

The LHC is a proton-proton collider currently have started
at CERN. With the designed
center-of-mass energy $14$ TeV and luminosity $L=10^{34}
cm^{-2}s^{-1}$, the LHC gives access to high energy frontier at TeV
scale and an opportunity to further improve the consistency test for
the CKM matrix. The production rates for heavy quark flavours will
be large at the LHC, and the $b\bar{b}$ production cross section
will be of the order 0.5 mb, providing as many as $0.5\times
10^{12}$ bottom events per year {\cite{Schopper2005}}. The heavy quark physics is one of
the main topics of LHC experiments.
Especially, LHCb detector is designed to
make precise studies on $CP$ asymmetries
and rare decays of b-hadron systems. The other
two experiments, ATLAS and CMS, are optimized for discovering new
physics and will complete most of their $B$ physics program within
the first few years {\cite{Schopper2005,Gouz2004}}.
Recently, the LHCb collaboration found clear evidence for
direct $CP$ violation in some three-body decay channels in
charmless decays of $B$ meson. Meanwhile,
large $CP$ violation is obtained in
$B^{\pm}\rightarrow \pi^{\pm}\pi^{+}\pi^{-}$ in the region
$0.6$ GeV${^2}$$<m^{2}_{\pi^{+}\pi^{-}low}<0.8$ GeV${^2}$ and $m^{2}_{\pi^{+}\pi^{-}high}>14$ GeV${^2}$. A zoom of the low $\pi^+\pi^-$ invariant mass from the $B^{\pm}\rightarrow \pi^{\pm}\pi^{+}\pi^{-}$ decay, showing
the region $0.6$ GeV${^2}$$<m^{2}_{\pi^{+}\pi^{-}low}<0.8$ GeV${^2}$ zone{\cite{LHC1301}}. Fortunately,
the experiments on $B_c$ mesons have been planned at LHCb. The predicted $CP$ violation for the decay processes we are considering
can be searched in the region of the invariant masses of $\pi^{+}\pi^{-}$
associated $\omega$ resonance on these experiments.

\begin{acknowledgments}
This work was supported by National Natural Science
Foundation of China (Project Numbers 11147003), Plan For Scientific Innovation Talent of Henan University of Technology
 (Project Number 2012CXRC17), the Key Project (Project Number 14A140001)
 for Science and Technology of the Education Department Henan Province,
the Fundamental Research Funds (Project Number 2014YWQN06) for the Henan Provincial Colleges and Universities,
and the Research Foundation of the young core teacher from Henan province.

\end{acknowledgments}


\section{APPENDIX: Related functions defined in the text}

In this paper, the related functions can be written as \cite{pqcd,pqcd1}\cite{Lcd:the two-body}\cite{prd81014022}:
\begin{eqnarray}\label{FLLe}
\mathcal {F}^{LL}_e&=&2\sqrt{\frac{2}{3}}C_Ff_B f_{P}\pi M_B^4\int_0^1dx_2\int_0^{\infty}b_1b_2db_1db_2    \nonumber\\
&\times&\phi_{D}(x_2,b_2)\{[(1-2r_{D})x_2+(r_D-2)r_b]                       \nonumber\\
&\times&\alpha_s(t_a)h_e(\alpha_e,\beta_a,b_1,b_2)S_t(x_2)\exp[-S_{ab}(t_a)]     \nonumber\\
&-&(r_D-2)r_D(x_1-1)\alpha_s(t_b)h_e(\alpha_e,\beta_b,b_2,b_1)        \nonumber\\
&\times &S_t(x_1)\exp[-S_{ab}(t_b)],
\end{eqnarray}
\begin{eqnarray}\label{FLRe}
\mathcal {F}^{LR}_e&=&-{F}^{LL}_e
\end{eqnarray}
\begin{eqnarray}\label{FSPe}
\mathcal {F}^{SP}_e&=&-4\sqrt{\frac{2}{3}}C_ff_Bf_{P}\pi M_B^4\int_0^1dx_2\int_0^{\infty}b_1b_2db_1db_2   \nonumber\\
&\times&\phi_{D}(x_2,b_2)\{[r_D(4r_b-x_2-1)-r_b+2]\alpha_s(t_a)     \nonumber\\
&\times&h_e(\alpha_e,\beta_a,b_1,b_2)S_t(x_2)\exp[-S_{ab}(t_a)]           \nonumber\\
&+&[r_D(2-4x_1)+x_1]\alpha_s(t_b)h_e(\alpha_e,\beta_b,b_2,b_1)    \nonumber\\
&\times&S_t(x_1)\exp[-S_{ab}(t_b)].
\end{eqnarray}
\begin{eqnarray}\label{MLLe}
\mathcal {M}_e^{LL}&=&\frac{8}{3}C_Ff_B\pi M_B^4\int_0^1dx_2dx_3\int_0^{\infty}b_2b_3db_2db_3   \nonumber\\
&\times&\phi_{D}(x_2,b_2)\phi_P^A(x_3)\{[r_D(1-x_1-x_2)+x_1                                                 \nonumber\\
&+&x_3-1] \alpha_s(t_c)h_e(\beta_c,\alpha_e,b_3,b_2)\exp[-S_{cd}(t_c)]                                  \nonumber\\
&-&[r_D(1-x_1-x_2)+2x_1+x_2-x_3-1]\alpha_s(t_d)                                                           \nonumber\\
&\times&h_e(\beta_d,\alpha_e,b_3,b_2)\exp[-S_{cd}(t_d)]\},
\end{eqnarray}
\begin{eqnarray}
\mathcal {M}_e^{LR}&=&\frac{8}{3}C_Ff_B\pi M_B^4 r_P(1+r_D)\int_0^1dx_2dx_3\int_0^{\infty}b_2b_3db_2db_3     \nonumber\\
&\times&\phi_{D}(x_2,b_2)\{[(x_1+x_3-1+r_D(2x_1+x_2+x_3-2))                                                                \nonumber\\
&\times&\phi_P^P(x_3)+(x_1+x_3-1+r_D(x_3-x_2))\phi_P^T(x_3)]                                                                \nonumber\\
&\times&\alpha_s(t_c)h_e(\beta_c,\alpha_e,b_3,b_2)\exp[-S_{cd}(t_c)]                                                       \nonumber\\
&- &[(x_1-x_3+r_D(2x_1+x_2-x_3-1))\phi_P^P(x_3)                                                                \nonumber\\
&+ &(x_3-x_1+r_D(x_3+x_2-1))\phi_P^T(x_3)]\alpha_s(t_d)                                                      \nonumber\\
&\times &h_e(\beta_d,\alpha_e,b_3,b_2)\exp[-S_{cd}(t_d)]\},
\end{eqnarray}
\begin{eqnarray}
\mathcal {M}_e^{SP}&=&\frac{8}{3}C_Ff_B\pi M_B^4\int_0^1dx_2dx_3\int_0^{\infty}b_2b_3db_2db_3           \nonumber\\
&\times&\phi_{D}(x_2,b_2)\phi_P^A(x_3)\{[r_D(x_1+x_2-1)-2x_1-x_2                                       \nonumber\\
&-&x_3+2]\alpha_s(t_c)h_e(\beta_c,\alpha_e,b_3,b_2)\exp[-S_{cd}(t_c)]                                    \nonumber\\
&-&[x_3-x_1-r_D(1-x_1-x_2)]\alpha_s(t_d)                                                                 \nonumber\\
&\times&h_e(\beta_d,\alpha_e,b_3,b_2)\exp[-S_{cd}(t_d)]\},
\end{eqnarray}
\begin{eqnarray}
\mathcal {F}_a^{LL}&=&\mathcal {F}_a^{LR}=-8C_Ff_B\pi M_B^4\int_0^1dx_2dx_3\int_0^{\infty}b_2b_3db_2db_3    \nonumber\\
&\times&\phi_{D}(x_2,b_2)\{[\phi_P^A(x_3)(x_3-2r_Dr_c)+r_P[\phi_P^P(x_3)                                      \nonumber\\
&\times&(2r_D(x_3+1)-r_c)+\phi_P^T(x_3)(r_c+2r_D(x_3-1))]]                                                     \nonumber\\
&\times&\alpha_s(t_e)h_e(\alpha_a,\beta_e,b_2,b_3)\exp[-S_{ef}(t_e)]S_t(x_3)                                   \nonumber\\
&-&[x_2\phi_P^A(x_3)+2r_Pr_D(x_2+1)\phi_P^P(x_3)]\alpha_s(t_f)                                             \nonumber\\
&\times&h_e(\alpha_a,\beta_f,b_3,b_2)\exp[-S_{ef}(t_f)]S_t(x_2)\},
\end{eqnarray}
\begin{eqnarray}
\mathcal {F}_a^{SP}&=&16C_Ff_B\pi M_B^4\int_0^1dx_2dx_3\int_0^{\infty}b_2b_3db_2db_3         \nonumber\\
&\times&\phi_{D}(x_2,b_2)\{[-\phi_P^A(x_3)(2r_D-r_c)+r_P[\phi_P^P(x_3)                    \nonumber\\
&\times&(4r_cr_D-x_3)+\phi_P^T(x_3)x_3]]\alpha_s(t_e)h_e(\alpha_a,\beta_e,b_2,b_3)       \nonumber\\
&\times&\exp[-S_{ef}(t_e)]S_t(x_3)-[x_2r_D\phi_P^A(x_3)+2r_P\phi_P^P(x_3)]                 \nonumber\\
&\times&\alpha_s(t_f)h_e(\alpha_a,\beta_f,b_3,b_2)\exp[-S_{ef}(t_f)]S_t(x_2)\};               \nonumber\\
\end{eqnarray}
\begin{eqnarray}
\mathcal {M}_a^{LL}&=&-\frac{8}{3}C_Ff_B\pi M_B^4
\int_0^1dx_2dx_3\int_0^{\infty}b_1b_2db_1db_2                         \nonumber\\
&\times&\phi_{D}(x_2,b_2)\{[\phi_P^A(x_3)(r_c-x_1+x_2)+r_Pr_D            \nonumber\\
&\times&[\phi_P^T(x_3)(x_2-x_3)+\phi_P^P(x_3)(4r_c-2x_1+x_2          \nonumber\\
&+&x_3)]]\alpha_s(t_g)h_e(\beta_g,\alpha_a,b_1,b_2)\exp[-S_{gh}(t_g)]    \nonumber\\
&+&[-\phi_P^A(x_3)(r_b+x_1+x_3-1)+r_Pr_D[(x_2-x_3)                         \nonumber\\
&\times&\phi_P^T(x_3)-\phi_P^P(x_3)(4r_b+2x_1+x_2+x_3-2)]]     \nonumber\\
&\times&\alpha_s(t_h)h_e(\beta_h,\alpha_a,b_1,b_2)\exp[-S_{gh}(t_h)]\},
\end{eqnarray}
\begin{eqnarray}
\mathcal {M}_a^{LR}&=&\frac{8}{3}C_Ff_B\pi M_B^4\int_0^1dx_2dx_3\int_0^{\infty}b_1b_2db_1db_2       \nonumber\\
&\times&\phi_{D}(x_2,b_2)\{[-\phi_P^A(x_3)r_D(r_c+x_1-x_2)+r_P                     \nonumber\\
&\times&[-\phi_P^T(x_3)(-r_c-x_1+x_3)+\phi_P^P(x_3)(r_c+x_1                        \nonumber\\
&-&x_3)]]\alpha_s(t_g)h_e(\beta_g,\alpha_a,b_1,b_2)\exp[-S_{gh}(t_g)]              \nonumber\\
&+&[-\phi_P^A(x_3)r_D(-r_b+x_1+x_2-1)+r_P[(-r_b                                    \nonumber\\
&+&x_1+x_3-1)(\phi_P^P(x_3)+\phi_P^T(x_3))]]\alpha_s(t_h)                           \nonumber\\
&\times&h_e(\beta_h,\alpha_a,b_1,b_2)\exp[-S_{gh}(t_h)]\},
\end{eqnarray}
\begin{eqnarray}\label{MSPa}
\mathcal {M}_a^{SP}&=&-\frac{8}{3}C_Ff_B\pi M_B^4\int_0^1dx_2dx_3\int_0^{\infty}b_1b_2db_1db_2                         \nonumber\\
&\times&\phi_{D}(x_2,b_2)\{[-\phi_P^A(x_3)(x_1-x_3-r_c)+r_Pr_D                                                          \nonumber\\
&\times&[-\phi_P^T(x_3)(x_2-x_3)+\phi_P^P(x_3)(4r_c-2x_1+x_2                        \nonumber\\
&+&x_3)]]\alpha_s(t_g)h_e(\beta_g,\alpha_a,b_1,b_2)\exp[-S_{gh}(t_g)]              \nonumber\\
&+&[-\phi_P^A(x_3)(r_b+x_1+x_2-1)+r_Pr_D[(-4r_b                                      \nonumber\\
&-&2x_1-x_2x_3+2)\phi_P^P(x_3)-(x_2-x_3)\phi_P^T(x_3))]]                            \nonumber\\
&\times&\alpha_s(t_h)h_e(\beta_h,\alpha_a,b_1,b_2)\exp[-S_{gh}(t_h)]\},
\end{eqnarray}
where $r_D=m_D/M_B$, $r_b=m_b/M_B$, $C_F=4/3$ is a color factor, $r_P=m^0_P/M_B$, with $m^0_P$ as the chiral mass of the pseudoscalar meson P.

For the $D_{(s)}$ meson wave function, we adopt the same model as of the B meson:
\begin{eqnarray}
\phi_{D_{(s)}}(x,b)&=&N_{D_{(s)}}[x(1-x)]^2        \nonumber\\
&&\exp\left(-\frac{x^2m_{D_{(s)}}^2}{2\omega_{D_{(s)}}^2}-\frac{1}{2}\omega_{D_{(s)}}^2b^2\right)
\end{eqnarray}
with  $\omega_{D}=0.6$ GeV.

We show here the functions $h_e$, coming from the Fourier transform of hard kernel.
\begin{eqnarray}
h_e(\alpha,\beta,b_1,b_2)&=&h_1(\alpha, b_1) \times h_2(\beta,b_1,b_2),\nonumber\\
h_1(\alpha, b_1)&=&\left\{\begin{array}{ll}K_0(\sqrt{\alpha} b_1),& \alpha>0\\
K_0(\sqrt{- \alpha} b_1),& \alpha<0
\end{array}
\right. \nonumber\\
h_2(\beta,b_1,b_2)&=&\left\{\begin{array}{ll}\theta(b_1-b_2)I_0(\sqrt{\beta}b_2)K_0(\sqrt{\beta}b_1)  \\
\theta(b_1-b_2)J_0(\sqrt{-\beta}b_2)K_0(i\sqrt{-\beta}b_1)
\end{array}
\right.  \nonumber\\
&+&\left\{\begin{array}{ll}+(b_1 \leftrightarrow b_2),       & \beta>0\\
+(b_1 \leftrightarrow b_2),& \beta<0
\end{array}
\right.
\end{eqnarray}
Where $J_0$ is the Bessel function and $K_0$, $I_0$ are modified Bessel function with $K_0(ix)=\frac{\pi}{2}(-N_0(x)+iJ_0(x))$. The hard scale t is chosen as the maximum of the virtuality of the internal momentum transition in the hard amplitudes, including $1/b_i(i=1,2,3)$:
\begin{eqnarray}
t_a&=&\mbox{max}\{{\sqrt{|\alpha_e|},\sqrt{|\beta_a|},1/b_1,1/b_2}\},\nonumber\\
t_b&=&\mbox{max}\{{\sqrt{|\alpha_e|},\sqrt{|\beta_b|},1/b_1,1/b_2}\},\nonumber\\
t_c&=&\mbox{max}\{{\sqrt{|\alpha_e|},\sqrt{|\beta_c|},1/b_2,1/b_3}\},\nonumber\\
t_d&=&\mbox{max}\{{\sqrt{|\alpha_e|},\sqrt{|\beta_d|},1/b_2,1/b_3}\},\nonumber\\
t_e&=&\mbox{max}\{{\sqrt{|\alpha_a|},\sqrt{|\beta_e|},1/b_2,1/b_3}\},\nonumber\\
t_f&=&\mbox{max}\{{\sqrt{|\alpha_a|},\sqrt{|\beta_f|},1/b_2,1/b_3}\},\nonumber\\
t_g&=&\mbox{max}\{{\sqrt{|\alpha_a|},\sqrt{|\beta_g|},1/b_1,1/b_2}\},\nonumber\\
t_h&=&\mbox{max}\{{\sqrt{|\alpha_a|},\sqrt{|\beta_h|},1/b_1,1/b_2}\},
\end{eqnarray}
where
\begin{eqnarray}
\alpha_e&=&(1-x_1-x_2)(x_1-r^2_D)M^2_B,\nonumber\\
\alpha_a&=&-x_2x_3(1-r^2_D)M^2_B,\nonumber\\
\beta_a&=&[r^2_b-x_2(1-r^2_D)]M^2_B,\nonumber\\
\beta_b&=&-(1-x_1)(x_1-r^2_D)]M^2_B,\nonumber\\
\beta_c&=&-(1-x_1-x_2)[1-x_1-x_3(1-r^2_D)]M^2_B,\nonumber\\
\beta_d&=&(1-x_1-x_2)[x_1-x_3-r^2_D(1-x_3)]M^2_B,\nonumber\\
\beta_e&=&[r^2_c-x_3-(1-x_3)r^2_D]M^2_B,\nonumber\\
\beta_f&=&-x_2(1-r^2_D)]M^2_B,\nonumber\\
\beta_g&=&[r^2_c-(x_1-x_3(1-r^2_D))(x_1-x_2)]M^2_B,\nonumber\\
\beta_h&=&[r^2_b-(1-x_1-x_3+x_3r^2_D)(1-x_1-x_2)]M^2_B,\nonumber\\
\end{eqnarray}

The $S_t$ re-sums the threshold logarithms $\ln^2x$ appearing in the
hard kernels to all orders and it has been parameterized as
\begin{eqnarray}
S_t(x)=\frac{2^{1+2c}\Gamma(3/2+c)}{\sqrt \pi
\Gamma(1+c)}[x(1-x)]^c,
\end{eqnarray}
with $c=0.4$. In the nonfactorizable contributions, $S_t(x)$ gives
a very small numerical effect to the amplitude~\cite{kurimoto}.
Therefore, we drop $S_t(x)$ in $F^{LL}_e$, $F^{SP}_e$, $F^{LL}_a$ and $F^{SP}_a$.

The Sudakov factors used in the text are defined by
\begin{eqnarray}
S_{ab}(t) &=& s\left(\frac{M_B}{\sqrt{2}}x_1, b_1\right)+s\left(\frac{M_B}{\sqrt{2}}x_2, b_2\right)  \nonumber \\
&+&\frac{5}{3}\int_{1/b_{1}}^{t}\frac{d \mu}{ \mu}\gamma_q (\mu)+2\int_{1/b_{2}}^{t}\frac{d \mu}{ \mu}\gamma_q (\mu)
\label{wp}
\end{eqnarray}
\begin{eqnarray}
S_{cd}(t) &=& s\left(\frac{M_B}{\sqrt{2}}x_1, b_2\right)+s\left(\frac{M_B}{\sqrt{2}}x_2, b_2\right)  \nonumber \\
&+& s\left(\frac{M_B}{\sqrt{2}}x_3, b_3\right)+s\left(\frac{M_B}{\sqrt{2}}(1-x_3), b_3\right)  \nonumber \\
&+&\frac{11}{3}\int_{1/b_{2}}^{t}\frac{d \mu}{ \mu}\gamma_q (\mu)+2\int_{1/b_{3}}^{t}\frac{d \mu}{ \mu}\gamma_q (\mu)
\label{Sc}
\end{eqnarray}
\begin{eqnarray}
S_{ef}(t) &=& s\left(\frac{M_B}{\sqrt{2}}x_2, b_2\right)+s\left(\frac{M_B}{\sqrt{2}}x_3, b_3\right)  \nonumber \\
&+&s\left(\frac{M_B}{\sqrt{2}}(1-x_3), b_3\right)+2\int_{1/b_{2}}^{t}\frac{d \mu}{ \mu}\gamma_q (\mu)  \nonumber \\
&+&2\int_{1/b_{3}}^{t}\frac{d \mu}{ \mu}\gamma_q (\mu)
\label{Se}
\end{eqnarray}
\begin{eqnarray}
S_{gh}(t) &=& s\left(\frac{M_B}{\sqrt{2}}x_1, b_1\right)+s\left(\frac{M_B}{\sqrt{2}}x_2, b_2\right)  \nonumber \\
&+& s\left(\frac{M_B}{\sqrt{2}}x_3, b_2\right)+s\left(\frac{M_B}{\sqrt{2}}(1-x_3), b_2\right)  \nonumber \\
&+&\frac{5}{3}\int_{1/b_{1}}^{t}\frac{d \mu}{ \mu}\gamma_q (\mu)+4\int_{1/b_{2}}^{t}\frac{d \mu}{ \mu}\gamma_q (\mu)
\label{ww}
\end{eqnarray}
$\gamma_q=-\alpha_s/\pi$ is the anomalous dimension of the quark. The explicit form for the  function
$s(Q,b)$ is:
\begin{eqnarray}
s(Q,b)&=&\int_{1/b}^{Q}\frac{d \mu}{\mu}[ln(\frac{Q}{\mu})A(\alpha_{s}(\mu))+B(\alpha_{s}(\mu))],\label{Sqb1}
\end{eqnarray}
where the anomlous dimensions $A$ two loops and $B$ to one loop are
\begin{eqnarray}
A&=&C_{F}\frac{\alpha}{\pi}+[\frac{67}{9}-\frac{\pi^2}{3}-\frac{10}{27}n_{f}+\frac{2}{3}\beta_{0}ln(\frac{e^{\gamma_E}}{2})](\frac{\alpha_{s}}{\pi})^2, \nonumber\\
B&=&\frac{2\alpha_s}{3\pi}ln(\frac{e^{2\gamma_{E}-1}}{2}),
\end{eqnarray}
with $C_F=4/3$ a color factor and $\gamma_{E}$ the Euler constant. The one-loop expression of the running coupling constant,
\begin{eqnarray}
\alpha_{\mu}&=&\frac{4\pi}{\beta_{0}ln(\mu^2/\Lambda^2)}
\end{eqnarray}
is substituted into Eq.(\ref{Sqb1}) with the coefficient $\beta_0=(33-2n_{f})/3$. $n_{f}$ is the number of the quark flavors.

Here, we specify the light-cone distribution amplitudes(LCDAs) for pseudoscalar and vector mesons. The twist-2 pseudoscalar meson distribution amplitude $\phi^A_p(P=\pi, K)$, and the twist-3 ones $\phi^P_p$ and $\phi^T_p$ have been parametrized as \cite{prd81014022},
\begin{eqnarray}
\phi^A_p(x)&=&\frac{f_P}{\sqrt{6}}3x(1-x)[1+a^{P}_{1}C^{3/2}_{1}(t)  \nonumber\\
&+& a^{P}_{2}C^{3/2}_{2}(t)+a^{P}_{4}C^{3/2}_{4}(t)]  \\
\phi^P_p(x)&=&\frac{f_P}{2\sqrt{6}}[1+(30\eta_{3}-\frac{5}{2}\rho^{2}_{P})C^{1/2}_2(t)  \nonumber\\
&-&3(\eta_3\omega_3+\frac{9}{20}\rho^2_P(1+6a^p_2))C^{1/2}_4(t)]  \\
\phi^T_p(x)&=&\frac{f_P}{2\sqrt{6}}(1-2x)[1+6(5\eta_{3}-\frac{1}{2}\eta_3\omega_3   \nonumber\\
&-&\frac{7}{20}\rho^2_P-\frac{3}{5}\rho^2_{P}a^P_2)(1-10x+10x^2)]
\end{eqnarray}
Where $t=2x-1$. For pseudoscalar mesons, we choose $\eta_{3}=0.015$ and $\omega_3=-3$. The mass ratio $\rho_{\pi (K)}=m_{\pi (K)}/m^{\pi (K)}_{0}$ and $\rho_{\eta_{q(s)}}=2m_{q(s)}/m_{qq(ss)}$, and the Gegenbauer polynomials $C^{\nu}_{n}(t)$ read
\begin{equation}
 \begin{array}{ll}
 C_2^{1/2} (t) = \frac{1}{2} (3t^2-1), & C_4^{1/2} (t) = \frac{1}{8}(+3-30t^2+35t^4),\\
 C_2^{3/2} (t) = \frac{3}{2} (5t^2-1), & C_4^{3/2} (t) = \frac{15}{8}(1-14t^2+21t^4),\\
 C_1^{3/2}(t)= 3t.
 \end{array}
 \end{equation}

\end{document}